\documentstyle[12pt]{article}

\newcommand{\be}{\begin{equation}}
\newcommand{\ee}{\end{equation}}
\newcommand{\beq}{\begin{eqnarray}}
\newcommand{\eeq}{\end{eqnarray}}
\newcommand{\ba}{\begin{array}{1}}
\newcommand{\ea}{\end{array}}
\newcommand{\bb}{\begin{thebibliography}}
\newcommand{\eb}{\end{thebibliography}}
\newcommand{\ci}[1]{\cite{#1}}
\newcommand{\bi}[1]{\bibitem{#1}}

\begin{document}
\title{
Probing the Deuteron Structure at small $NN$ Distances  \\
by Antiproton-Deuteron Annihilation
\footnote{supported by
the Russia Foundation for the fundamental Research, N 96-02-16126}
}
\author{
 O.Yu.Denisov \thanks{On leave of absence from JINR, Dubna, Russia}\\
 {\it Istituto di Fisica, Universit\`a di Torino and INFN,} \\
 {\it Sez. di Torino, Torino, Italy } \\
 \and
   S. D. Kuksa, G. I. Lykasov  \\
 {\it Joint Institute for Nuclear Research}\\
 {\it 141980, Dubna, Moscow Region, Russia}
}

\maketitle

\newpage

\begin{abstract}

 The annihilation $\bar p d\rightarrow 2\pi^-\pi^+p$ at rest is
 analyzed. Assuming that the deuteron wave function can be represented
 by the Fock's column consisting of different possible states
 ($~NN, NN^*,\Delta\Delta, NN\pi$,...)
 some enhancement in the distribution
 over the invariant mass of the $\pi^-\pi^-$ system
 in the mass region $~1.4-1.5(GeV/c^2)$ is predicted.
 This can be caused by the possible existence of a $\Delta-\Delta$
 component in deuteron. Using this prediction and new experimental
 data of the OBELIX collaboration (LEAR, CERN) one can estimate the upper
 limit for this exotic state of deuteron.

\end{abstract}

\vspace{1cm}


\newpage

 Over the past decades the deuteron structure at short distances
 has been investigated rather intensively. There are many theoretical
 approaches trying to study this problem. For example, within the
 quark model picture the admixture of 6q-components has been
 suggested \ci{1,2}. This corresponds to an admixture of baryon
 resonances within the asymptotic basis of baryonic states
 ($N(938), \Delta(1232), N^*(1440),...$) \ci{3,4}. New
 information about the existence of, for example, $N N^*(1440)$
 in a deuteron can be obtained by studying subthreshold antiproton
 production in $pd$ and $dd$ reactions \ci{5}. It was shown that the
 probability of this component in a deuteron can be less than
 $0.5\%$. Attempts
 to detect baryon resonances in a deuteron were made by many
 experimental groups \ci{exp1,exp2} and \ci{exp3,exp4}. As a rule,
 pion, photon, neutrino and proton beams of intermediate energy
 together with fixed target setups were used for this purpose.
 These groups have tried to demonstrate the presence of $N^*$ in
 nuclei by shaking loose an $N^*$ that preexists in the nucleus
 acting as a spectator in some reactions. The main problem for such
 experiments was to distinguish between the signal from the internal
 $N^*$ and the signal from another states produced by the conventional
 way, e.g. via rescattering. The experimental results are given as upper
 limits, ranging in $(0.4~-~0.9)\%$.

 We propose alternatively to
 explore antiproton annihilation on the deuteron to provide
 new information on the possible existence of the above mentioned
 states in the deuteron. Here we are intend well known
 measurements of $\bar p d$ annihilation at rest performed by the
 OBELIX collaboration (see e.g. \ci{6}).
 New data obtained by this experiment \ci{7} show the very interesting
 behavior of the $\pi^-\pi^-$ system invariant mass distribution
 in reaction $\bar p d\rightarrow 2\pi^-\pi^+p$. In this paper we
 concentrate on detailed analysis of this annihilation
 channel. We show also that some enhancement
 in the spectrum over the invariant mass of two $\pi^-$-mesons
 observed by OBELIX can be explained
 by the possible existence of a $\Delta-\Delta$ component in the deuteron.

% \begin{figure}[ht]
% \psfull
% \begin{center}
% \mbox{
% \epsfig{file=diag_lyk1.eps,height=3.cm,width=5.cm}}
% \end{center}
% \caption[diag1]{Impulse approximation for the
% $\bar p d\rightarrow 2\pi^-\pi^+ p$ annihilation.}
% \label{diag1}
% \end{figure}

 The general expression for the differential cross section
 $d\sigma/dm_{2\pi}$ for the above mentioned annihilation reaction
 in flight within the framework of the impulse approximation, see Fig.1,
 can be written in the following form:
 \begin{eqnarray}
 \frac{d\sigma}{dm_{2\pi}}=\frac{2m_{2\pi}}{{(2\pi)}^8
 2\lambda^{1/2}(s,m^2,M_d^2)}
 \int ds_{3\pi}R_2(s,m^2,M^2_d)R_3(s_{3\pi},m^2_n,m^2)          \\ \nonumber
 {\mid T_{\bar p d\rightarrow 2\pi^-\pi^+ p }(s,s_{3\pi},t,t_1)
 \mid }^2
 \label{1}
 \end{eqnarray}
 where the following notation is introduced: $m_{2\pi}$ is the
 invariant mass of two pions; $\lambda^{1/2}(x,y,z)=
 (x-(\sqrt{y}-\sqrt{z})^2)^{1/2}(x-(\sqrt{y}+\sqrt{z})^2)^{1/2}$;
 $s$ is the square initial energy in the $\bar p-d$ c.m.s.;
 $s_{3\pi}$ is the square invariant mass of three pions;
 $m,M_d$ are the masses of the nucleon (antinucleon) and the deuteron,
 respectively; $m_n$ is the mass of the neutron inside the deuteron
 which, in principle, is off-shell; $R_2(s,m^2,M^2_d)$ is the
 two-particle relativistic invariant phase-space for the binary process
 $\bar p + d\rightarrow 3\pi + p$ when in the final state there are
 a proton and a particle with the invariant mass of three pions;
 $R_3(s_{3\pi},m^2_n,m^2)$ is the three-particle relativistic invariant
 phase-space for annihilation of the initial antiproton on the
 intra-deuteron neutron $\bar p n\rightarrow 2\pi^-\pi^+$;
 $T_{\bar p d\rightarrow 3\pi p}(s,s_{3\pi},t,t_1)$ is the amplitude
 of the process $\bar p d\rightarrow 2\pi^-\pi^+ p$, $t=(p_d-p_p)^2$,
 where $p_d, p_p$ are the four-momenta of the initial deuteron and
 final proton, respectively; $t_1=(p_{\bar p}-p_\pi)^2$, where
 $p_{\bar p}, p_\pi$ are the four-momenta of the $\bar p$ and of the
 final pions, respectively.

 Within the framework of the impulse approximation, see Fig.1, this
 amplitude can be written in the factorized form:
 \begin{eqnarray}
 T_{\bar p d\rightarrow 3\pi p}(s,s_{3\pi},s_{2\pi},t,t_1)=
 \phi_d({\bf p}_p(s,t))f_{\bar p n\rightarrow 2\pi^-\pi^+}
 (s_{3\pi},s_{2\pi},t_1)
 \label{g2}
\end{eqnarray}
 where $\phi_d$ and $f_{\bar p n\rightarrow 2\pi^-\pi^+}$ are the
 deuteron wave function and the amplitude of the annihilation process
 $\bar p n\rightarrow 2\pi^-\pi^+$; ${\bf p}_p$ is the three-momentum
 of the final proton. Now, inserting now expressions for $R_2,R_3$ \ci{8}
 and $T_{\bar p d\rightarrow 3\pi p}$ given by eq.(\ref{g2}) into
 eq.(1) one can get the following form for the distribution
 over the invariant mass of the $\pi^+\pi^-$-pair:
 \begin{eqnarray}
 \frac{d\sigma}{dm_{\pi^+\pi^-}}=\frac{2m_{\pi^+\pi^-}\pi^3}
 {2(2\pi)^8 2\lambda(s,m^2,M^2_d)}\frac{1}
 {2\lambda^{1/2}(s_{\pi^+\pi^-},\mu^2,\mu^2)}
 \int ds_{3\pi}\int\frac{dt}{2\lambda^{1/2}(s_{3\pi,t,m^2})}  \\ \nonumber
 {\mid \phi^2_d(p(s(t))\mid}^2
 \int dt_1\int ds_{2\pi^-}
 {\mid f_{\bar p n\rightarrow 2\pi^-\pi^+}(s_{3\pi},s_{2\pi},t_1)
 \mid}^2
 \label{g3}
 \end{eqnarray}
 A similar expression can be written for the distribution over
 the invariant mass of the $\pi^-\pi^-$-pair:
 \begin{eqnarray}
 \frac{d\sigma^{1}}{dm_{\pi^-\pi^-}}=\frac{2m_{2\pi^-}\pi^3}
 {2(2\pi)^8 2\lambda(s,m^2,M^2_d)}\frac{1}
 {2\lambda^{1/2}(s_{2\pi^-},\mu^2,\mu^2)}
 \int ds_{3\pi}\int\frac{dt}{2\lambda^{1/2}(s_{3\pi},t,m^2)}   \\ \nonumber
 {\mid \phi^2_d(p(s(t))) \mid}^2
 \int dt_1\int ds_{\pi^+\pi^-}
 {\mid f_{\bar p n\rightarrow 2\pi^-\pi^+}(s_{3\pi},s_{2\pi},t_1)
 \mid}^2
 \label{g4}
 \end{eqnarray}
 where $s_{2\pi}=m^2_{2\pi}$ is the square invariant mass of
 $2\pi^-$, $\mu$ is the pion mass.
 For calculation of the spectra given by eqs. (3), (4) the
 amplitude of the annihilation $\bar p n\rightarrow 2\pi^-\pi^+$
 should be known.

 Actually, the question arises how to calculate the amplitude
 $f_{\bar p n\rightarrow 2\pi^-\pi^+}$. It was parametrized by a
 form that resulted in a satisfactory description of the experimental
 data for $dN/dm_{\pi^+\pi^-}$. According to the OBELIX
 data \ci{7} the spectrum $dN/dm_{\pi^+\pi^-}$ has a resonance-like
 shape, see Fig.2, corresponding to the possible production of $\rho$-meson
 at $m_{\pi^+\pi^-}\simeq 0.77 (GeV/c^2)$ and $f_2$-meson at
 $m_{\pi^+\pi^-}=1.26(GeV/c^2)$. The dependence of
 ${\mid f_{\bar p n\rightarrow 2\pi^-\pi^+}(s_{3\pi},s_{2\pi},t_1)\mid}^2$
 on $s_{\bar p n}=s_{3\pi}$ can be found using the model of the Reggeized
 meson exchange in the $t$-channel of the reaction
 $\pi^+ n\rightarrow \pi^+\pi^- p$,
 where $t > 0.$ and
 $t=(p_n - p_p)^2=(p_n + p_{\bar p})^2\equiv s_{\bar p n}=s_{3\pi}$.
 The $t_1$-dependence of
 ${\mid f_{\bar p n\rightarrow 2\pi^-\pi^+}(s_{3\pi},s_{2\pi},t_1)\mid}^2$
 can be found using the usual one-baryon exchange model of the reaction
 $\bar p n\rightarrow 2\pi^-\pi^+$. Therefore, the square of the amplitude
 $f_{\bar p n\rightarrow 2\pi^-\pi^+}(s_{3\pi},s_{2\pi},t_1)$ has been
 parametrized by the following form:
 \begin{eqnarray}
 {\mid f_{\bar p n\rightarrow 2\pi^-\pi^+}(s_{3\pi},s_{2\pi},t_1)\mid}^2=
 B^2(s_{\pi^+\pi^-})\frac{s_{3\pi}}{(s_{3\pi}+\mu^2)^2}F^2_1(s_{3\pi})
 \frac{\mid t_1\mid}{(\mid t_1\mid +m^2)^2}F^2_2(t_1)
 \label{g5}
 \end{eqnarray}
 where $F_1(s_{3\pi})$ is the pion form-factor corresponding to the
 $NN\pi$-vertex by the Reggeized meson-exchange of $\pi^+ n\rightarrow
 \pi^+\pi^-p$ process, it can be parametrized by the usual form
 $F_1(s_{3\pi})=exp(-s_{3\pi}R^2_\pi)$ where the parameter 
 $R^2_\pi\simeq 2GeV^{-2}$ \ci{9} is used; $F_2(t_1)$ is the baryon
 form-factor corresponding to the $NN\pi$-vertex by the one-baryon exchange
 in $s$-channel of the reaction $\bar p n\rightarrow 2\pi^-\pi^+$,
 it was taken in the form
 $F_2(t_1)=\Lambda^2_B/(\mid t_1\mid + \Lambda^2_B)$
 where $\Lambda_B=0.4-0.5(GeV/c)^2$ \ci{10};
 $B(s_{\pi^+\pi^-})$ is the function corresponding to the resonance
 and nonresonance mechanisms of the annihilation process
 $\bar p n\rightarrow 2\pi^-\pi^+$, it was parametrized by the
 following form:
 \begin{eqnarray}
 B(s_{\pi^+\pi^-})=\frac{w_{\rho}}{\sqrt{s_{\pi^+\pi^-}}-m_\rho
 -i\Gamma_\rho/2} + \frac{w_{f_2}}{\sqrt{s_{\pi^+\pi^-}}-m_{f_2}
 -i\Gamma_{f_2}/2} + iw_0
 \label{g6}
  \end{eqnarray}
 Really, the form (\ref{g5}) is some parametrization of the square
 of the amplitude of the process $\bar p n\rightarrow 2\pi^-\pi^+$,
 and application of the Reggeized one-boson exchange model can
 be considered only as a prompt to find the parametrization
 form (\ref{g5}).

 Now, inserting eq.(\ref{g5}) into eq.(3) one can calculate the
 spectrum  $d\sigma/dm_{\pi^+\pi^-}$. Experimental measurements
 of this spectrum were performed at rest and presented in \ci{7}. 
 Therefore, the spectrum given by eq.(3) was
 calculated at a very small initial antiproton energy satisfying to the
 experimental resolution and normalized by the total number of events, $N$,
 corresponding to experimental measurements. This spectrum
 $dN/dm_{\pi^+\pi^-}$ and the experimental data are presented
 in Fig.2, the parameters $w_\rho, w_{f_2}, w_0$ and $R^2_ \pi$
 were found by fitting the
 OBELIX  experimental data \ci{7} for
 $dN/dm_{\pi^+\pi^-}$,
 they are $w_\rho=1.2, w_{f_2}=1.35, w_0=35.0$ and
 $R^2_\pi\simeq 2.(FeV/c)^{-2}$. 

% \begin{figure}[ht]
% \psfull
% \begin{center}
% \mbox{
% \epsfig{file=pipan.eps,height=7.cm,width=5.cm}}
% \end{center}
% \caption[pipsp]{The $\pi^+\pi^-$ invariant mass spectrum.}
% \label{pipsp}
% \end{figure}

 Now, inserting the same form for
 ${\mid f_{\bar p n\rightarrow 2\pi^-\pi^+}(s_{3\pi},s_{2\pi},t_1)\mid}^2$
 given by eq.(\ref{g5}) into eq.(4)
 one can calculate the spectrum $dN/dm_{2\pi^-}$ which is presented in Fig.3
 (the dashed curve) together with the experimental data \ci{7}.

% \begin{figure}[ht]
% \psfull
% \begin{center}
% \mbox{
% \epsfig{file=piman.eps,height=7.cm,width=5.cm}}
% \end{center}
% \caption[pimsp]{The invariant mass spectrum for $2\pi^-$-
% mesons. The dashed line corresponds to the conventional deuteron
% consisting of a proton and a neutron. The solid line corresponds to the
% case when the $\Delta\Delta$ component in the deuteron is taken
% into account.}
% \label{pimsp}
% \end{figure}

 Now, consider the problem related to the possible existence of baryon
 resonances in a deuteron. There is a point of view \ci{frank,shap}
 that a deuteron wave function (d.w.f.) can be considered a Fock
 column the lines of which are $NN, \Delta\Delta, NN^*, NN\pi$, etc.,
 components. There are theoretical models within the framework of which
 the d.w.f. is considered taking into account these baryon resonances in
 a deuteron. We used one of these models \ci{1,2} in order to estimate
 the upper limit of the probability for a $\Delta\Delta$-component to exist
 in the deuteron. The possible mechanism of the annihilation
 $\bar p d\rightarrow 3\pi p$ is depicted in Fig.4.

% \begin{figure}[ht]
% \psfull
% \begin{center}
% \mbox{
% \epsfig{file=diag_lyk2.eps,height=3.cm,width=5.cm}}
% \end{center}
% \caption[diag2]{Graph for $\bar p d\rightarrow 2\pi^-\pi^+ p$
% annihilation assuming the existence of $\Delta^{++},\Delta^-$
% isobars in the deuteron.}
% \label{diag2}
% \end{figure}

 The contribution to the
 spectrum $d\sigma/dm_{2\pi^-}$ corresponding to this graph of Fig.4
 can be presented as:
 \begin{eqnarray}
 \frac{d\sigma^{\Delta\Delta}}{dm_{2\pi^-}}=\frac{2m_{2\pi^-}}{{(2\pi)}^8
 2\lambda^{1/2}(s,m^2,M_d^2)}
 \int ds_{\pi^+p}R_2(s,s_{2\pi^-},s_{\pi+p})R_2(s_{2\pi^-},\mu^2,\mu^2)   \\ \nonumber
 R_2(s_{\pi^+p},m^2,\mu^2)
 {\mid T_{\bar p d\rightarrow 2\pi^-\pi^+ p }(s,s_{2\pi^-},s_{\pi^+p},t_1)
 \mid }^2
 \label{g7}
 \end{eqnarray}
 The amplitude
 $T_{\bar p d\rightarrow 2\pi^-\pi^+ p }(s_{2\pi^-},s_{\pi^+p},t_1)$
 corresponding to the graph of Fig.4 is written in the following
 form:
 \begin{eqnarray}
 T_{\bar p d\rightarrow 2\pi^-\pi^+ p }(s_{2\pi^-},s_{\pi^+p},t_1)=
 \phi^{\Delta\Delta}_d(p_\Delta)f_{\bar p \Delta^-\rightarrow 2\pi^-}
 (t_1)f_{\Delta^{++}\rightarrow \pi^+p}(s_{\pi+p})
 \label{g8}
 \end{eqnarray}
 where $\phi^{\Delta\Delta}_d(p_\Delta)$ is the part of the d.w.f.
 corresponding to the $\Delta\Delta$ component of Fock's column;
 $f_{\bar p \Delta^-\rightarrow 2\pi^-}$ is the amplitude of the
 process $\bar p \Delta^-\rightarrow 2\pi^-$;
 $f_{\Delta^{++}\rightarrow \pi^+p}(s_{\pi+p})$ is the function
 corresponding to the down vertex
 of Fig.4, e.g., the decay of $\Delta^{++}\rightarrow \pi^+p$,
 and depending on the square of the invariant mass of the
 $\pi^+ p$ system $s_{\pi^+p}$ .
 The form of $\mid f_{\bar p \Delta^-\rightarrow 2\pi^-}\mid^2$
 can be found using the one-baryon exchange model, e.g.:
 \begin{eqnarray}
 \mid f_{\bar p \Delta^-\rightarrow 2\pi^-}\mid^2(t_1)=
 \frac{\mid t_1\mid}{(\mid t_1\mid + m^2)^2}F^2_2(t_1)
 \label{g9}
  \end{eqnarray}
 The function $f_{\Delta^{++}\rightarrow \pi^+p}(s_{\pi+p})$ has
 the Breit-Wigner form and can be written as the following:
 \begin{eqnarray}
 \mid f_{\Delta^{++}\rightarrow \pi^+p}(s_{\pi+p})\mid^2=
 \frac{C}{(\sqrt{s_{\pi^+p}}-m_\Delta)^2+\Gamma^2_\Delta/4}
 \label{g10}
 \end{eqnarray}
 where $m_\Delta$ is the $\Delta$-isobar mass, $\Gamma_\Delta$ is its
 width, $C$ is some constant. The form of the d.w.f.
 $\phi^{\Delta\Delta}_d(p_\Delta)$ taking into account the possible
 existence of a $\Delta\Delta$-component in the deuteron was taken from
 \ci{1,2} obtained within the framework of the quark model
 and had the following form:
 \begin{eqnarray}
 \phi^{\Delta\Delta}_d(p_\Delta)=\sqrt{s_\Delta}
 (\frac{4b^2}{\sqrt{\pi}(3/2)^{3/2}})^{1/2} exp(-b^2p^2_\Delta/3)
 \label{g11}
 \end{eqnarray}
 where $s_\Delta$ is the so-called spectroscopic factor \ci{1,2};
 $b$ is the slope parameter \ci{1,2}.
 Now, inserting $R_2(s,s_{2\pi^-},s_{\pi^+p}), R_2(s_{2\pi^-},\mu^2,\mu^2),
 R_2(s_{\pi^+p},m^2,\mu^2)$ and
 eqs.(\ref{g8}-\ref{g11}) into eq.(7) one can get the expression for
 the contribution $d\sigma^{\Delta\Delta}/dm_{2\pi^-}$:
 \begin{eqnarray}
 \frac{d\sigma^{\Delta\Delta}}{dm_{2\pi^-}}=\frac{2m_{2\pi^-}\pi^3}
 {(2\pi)^8 2\lambda(s,m^2,M_d^2)2\lambda^{1/2}(s_{2\pi^-},\mu^2,\mu^2)}
 \int ds_{\pi^+p}\frac{\lambda^{1/2}(s_{2\pi^-},m^2,\mu^2)}
 {2s_{\pi^+p}}                                              \\ \nonumber
 \frac{C}{(\sqrt{s_{\pi^+p}}-m_\Delta)^2+\Gamma^2_{\Delta}/4}
 \int dt\mid\phi^{\Delta\Delta}_d(p^2_{\Delta}\mid^2\int dt_1
 \mid f_{\bar p \Delta^-\rightarrow 2\pi^-}(t_1)\mid^2
 \label{g12}
 \end{eqnarray}

 Then, the total spectrum $d\sigma/dm_{2\pi^-}$ within the framework
 of the impulse approximation can be obtained as the incoherent sum of
 $d\sigma^{(1)}/dm_{2\pi^-}$ and $d\sigma^{\Delta\Delta}/dm_{2\pi^-}$,
 e.g.:
 \begin{eqnarray}
 \frac{d\sigma}{dm_{2\pi^-}}=
 \frac{d\sigma^{1}}{dm_{2\pi^-}} +
 \frac{d\sigma^{\Delta\Delta}}{dm_{2\pi^-}}
 \label{l12}
 \end{eqnarray}
 The corresponding spectrum $dN/dm_{2\pi^-}$ 
 calculated in the same manner as $dN/dm_{\pi^+\pi^-}$ and the OBELIX
 experimental data \ci{7} are
 presented in Fig.3 (solid line). By the calculation of
 $d\sigma^{\Delta\Delta}/dm_{2\pi^-}$ the joint parameter value
 $s_\Delta*C^2$ has been found  fitting  the experimental data
 for $dN/dm_{2\pi^-}$ \ci{7}. 
 According to our calculations the spectrum
 $d\sigma^{\Delta\Delta}/dm_{2\pi^-}$ may contribute
 to the total spectrum at $m_{2\pi^-}=1.4-1.5 (GeV/c^2)$. Therefore, 
 the small enhancement in
 $dN/dm_{2\pi^-}$ observed experimentally at
 $m_{2\pi^-}=1.4-1.5 (GeV/c^2)$, see Fig.3, can be interpreted as the
 contribution of $d\sigma^{\Delta\Delta}/dm_{2\pi^-}$.

 At this point the question arises about the contribution of corrections
 to the impulse approximation of the annihilation process
 $\bar p d\rightarrow 2\pi^-\pi^+ p$, see Fig.1. As for this possible
 correction, the two-step mechanism \ci{lev} can contribute to the spectrum
 $d\sigma/dm_{2\pi}$.
 Our estimations and the results of ref.\ci{lev} showed that this contribution
 is very small for the invariant mass value of
 $\pi^+p$ $m_{\pi^+p} < 1.1-1.15 (GeV/c^2)$ tha corresponds to
 $m_{2\pi^-} > 1.3-1.35 (GeV/c^2)$ and can't give any enhancement
 at $m_{2\pi^-}=1.4-1.5 (GeV/c^2)$ to the spectrum $dN/dm_{2\pi^-}$,
 when according to \ci{7} the momentum of the proton $p~\geq~400~MeV/c$.
 In principle, the two-step mechanism can give some sizeable contribution
 to this spectrum at $m_{2\pi^-}\sim 0.8-1.1 (GeV/c^2)$. However, it is
 very difficult to calculate it accurately because there is
 a large uncertainty related to the off-shellness of the virtual meson in
 the intermediate state \ci{lyk,lrc}.
 Thus, according to the above this $\Delta\Delta$ component
 can contribute in the high mass kinematical region
 where the two-step mechanism of the annihilation process
 $\bar p d\rightarrow 2\pi^-\pi^+ p$ can be neglected.

 So, let us assume that a small enrichment in the $\pi^-~\pi^-~$ system
 invariant mass distribution in the reaction $\bar p d\rightarrow 2\pi^-\pi^+ p$
 is a manifistation of the $\Delta \Delta$ component of the deuteron.

% \begin{figure}[ht]
% \psfull
% \begin{center}
% \mbox{
% \epsfig{file=del_lyk.eps,height=6.cm,width=10.8cm}}
% \end{center}
% \caption[3ip]{
% \baselineskip=8pt
% (a)Scatter plot for the invariant mass of the $\pi^-\pi^-$
% system versus the invariant mass of the $\pi^+p$ system for the
% reaction $\bar p d\rightarrow 2\pi^-\pi^+p$. (b) The scatter
% plot projection on to the vertical axis. The hatched diagram corresponds
% to the events with the invariant mass of the $\pi^+p$ system satisfying
% the following condition:
% $\left| M_{\pi^+p}~-~M_{\Delta^{++}}\right|~\leq~120~MeV.$}
% \label{3ip}
% \end{figure}

 In this case, due to the lack of statistics, we can
 only obtain an upper limit for the process of Fig.4.
 Fig.5(a) shows the scatter plot for the invariant
 mass of the $\pi^-\pi^-$ system versus the invariant mass of the
 $\pi^+p$ for $\bar p d\rightarrow 2\pi^-\pi^+p$ annihilation
 via the channel of Fig.4.
 There is a some small enrichment in the upper part of the $\pi^+p$
 system in the $\Delta^{++}(1232)$ region.
 In Fig.5(b),  the invariant mass distribution of the $\pi^-\pi^-$
 system in this reaction is shown. The solid line corresponds
 to all the selected events. The hatched histogram corresponds
 to the events with the invariant mass of the $\pi^+p$ system
 satisfying the following condition:
 $\left|M_{\pi^+p}~-~M_{\Delta^{++}}\right|~\leq~120~MeV$.
 As one can see, there is a small bump in the high mass part of
 the $\pi^-\pi^-$ invariant mass distribution, but this bump has a
 small statistical significance, so we can only estimate the upper
 limit for the reaction of Fig.4. To do this the distribution
 in Fig.5(b) was approximated at a first stage by a second order
 polynomial function. The parameters of this polynomial function were
 fixed and the experimental invariant mass distribution of the
 $\pi^-\pi^-$ system was fitted at the second stage by the function:
 \beq
 F~=~A_1~\times~BW(\Gamma_{\pi^-\pi^-},M_{\pi^-\pi^-},m)~
 +~A_2~\times~\sum_{1}^{3}~b_i~\times~m^{i-1}
 \label{og1}
 \eeq
 where $BW(\Gamma_{\pi^-\pi^-},M_{\pi^-\pi^-},m)$ is the Breit-Wigner
 function. During the fit, the parameters $A_1,A_2,M_{\pi^-\pi^-}$
 were free, the parameters of the polynomial function were fixed from
 the first stage of approximation and the $\Gamma_{\pi^-\pi^-}$
 parameter was fixed at a value of $60~ MeV$ according to 
 theoretical prediction. Using the results of this fit,
 the upper limit on the reaction of Fig.4 was estimated to be:
 \beq
 Y_{\bar p~(\Delta^-\Delta^{++})\rightarrow 2\pi^-\pi^+p}~
 \leq~6.5~\times~10^{-5}
 \label{og2}
 \eeq
 with a $90\%$ confidence level.

 Finally, let us suppose that the branching ratio of the reaction
 $\bar p\Delta^-~\rightarrow ~2\pi^-$ is approximately the same
 as for $pN~\rightarrow ~2\pi$, that is $~\simeq ~5.~\times~10^{-3}$
 \ci{dos}. Using this assumption and the upper limit (\ref{og2}) we
 can roughly estimate the probability to find the deuteron as the
 $\Delta\Delta$ configuration. This probability is

                  $$P_{\Delta\Delta}~\leq~1.\%$$
 Note, that in the near future a
 factor of 10 higher statistics will be available for analysis from the OBELIX
 collaboration.

 Let us now make a short conclusion on our investigation.
 Some enrichment of the scatter plot for $m_{2\pi^-}$ versus
 $m_{\pi+ p}$ for the $\bar p d\rightarrow 2\pi^-\pi^+p$ process
 and the corresponding small enhancement in the spectrum
 $dN/dm_{2\pi^-}$ at $m_{2\pi^-}=1.4-1.5(GeV/c^2)$ observed
 experimentally \ci{7} can be caused by the possible existence
 of a $\Delta-\Delta$ component in the deuteron. The analysis
 of this phenomenon shows that it is difficult to calculate
 the absolute value of its probability. But the theoretical
 prediction about the width value of this enhancement allows
 us to estimate the upper limit for the $\Delta-\Delta$ exotic
 state of the deuteron.

 The authors wish to thank V.G.Neudatchin, I.T.Obuchovsky, I.Chuvilsky
 and also M.P.Bussa for valuable hints and inspiring discussions.

 \end{document}